\documentstyle[preprint,aps]{revtex}
\tightenlines 
\begin{document}
\title
{Adaptive Ising Model and  Bacterial Chemotactic Receptor Network}
\author{Yu Shi\cite{e}
} 
\address{Cavendish Laboratory, University of Cambridge, Cambridge CB3 0HE,
 United Kingdom}
\draft
\maketitle
\begin{abstract}
We present  a so-called adaptive Ising model (AIM)
 to provide  a unifying explanation for
 sensitivity and perfect adaptation in bacterial
 chemotactic signalling, based
on  coupling among receptor dimers.
In an AIM, an external  field, representing ligand binding, 
is  randomly applied 
to a fraction  of spins, representing the states
 of the receptor dimers,  and  
there is   a delayed negative feedback from the spin value on the  local 
 field.  This  model is solved in an adiabatic approach.
If the feedback is slow and weak  enough,  as indeed in 
chemotactic signalling,  
the system evolves through quasi-equilibrium states and 
the ``magnetization'', representing the signal, 
always
attenuates towards  zero and is always sensitive to a  subsequent stimulus.
\end{abstract}


As a prototypical system of cell signalling,  
bacterial chemotaxis is extensively studied in biology. 
  A bacterium,  such as 
{\it Escherichia coli},
swims  under the  control  of  several flagellar motors.
When the motors rotate counterclockwisely, the bacterium runs forward;
when the motors rotate clockwisely, it tumbles.  
The motors switch between these 
 two modes of rotation with the probability ratio 
 determined, through a signalling pathway,  by   the change 
of the concentration of the ambient chemical.
Therefore  the bacterium  performs a biased
random walk towards  higher concentration of 
an attractant or lower concentration of a repellent.
The signalling pathway  (Fig. 1) is as follows \cite{stock}.
The chemoeffector molecule ligands   bind
 to transmembrane receptor dimers, 
which are
 coupled by two proteins CheW to  two cytoplasmic histidine kinases CheA.
 CheA is autophosphorylated with the rate
greatly enhanced by the receptor, hence  attractant binding  causes the 
receptor dimer to undergo  a 
conformational  change which  leads to a decrease of 
autophosphorylation rate of CheA.
CheA transfers  phosphorylation group to two regulators CheB and CheY.
Phospho-CheY directly modulates the motors.
On the other hand, 
 Phospho-CheB mediates demethylation of the
receptor, while another regulator
CheR promotes  methylation. Attractant binding also
makes the receptor a  better substrate for CheR. 
Since methylation increases the autophosphorylation rate of CheA,
Phospho-CheB and CheR provide  a negative feedback.
The basic structure of the information 
loop is: the ligand binding is the input,
the  activity change of CheA is the output, and there is a negative feedback
from the output on the input.

A crucial feature of chemotaxis is its sensitivity:
as little as a single molecule can trigger a detectable motor response
\cite{block}.  Another crucial feature  is
adaptation:  after an initial sensitive response, the tumbling 
rate returns to the pre-stimulus level.
A noteworthy fact is that 
there are about $2000$ chemotactic  receptor dimers 
clustering at a  pole of the cell, furnishing a detecting ``nose''.
 Is the sensitivity  enhanced by clustering \cite{pb}?
The possibility that ligand binding of one receptor can change
the activities of 
more  receptors  was  considered \cite{bray}.
A  biological principle may be formulated:
{\em An attribute that exists most probably confers advantages over
 possible alternatives, especially if the latter have some 
apparent merit} \cite{shi}.
This principle and various  
experimental findings  led 
to    a  cooperative model based on coupling 
among receptor dimers  \cite{shi}.
 This  model is equivalent to an Ising
model in a bimodally distributed field, and  provides an arbitrarily sensitive 
 initial response,
 by choosing an appropriate
value of a parameter comparing the coupling with the noise. 
In this theoretical framework, 
 adaptation is  achieved by a  counteracting effect due to the   negative 
feedback and 
 mapping to an opposite induced field 
in the Ising model. The theoretical results  are in good agreement with,
say, a recent experiment \cite{khan}. 

More investigations  need to be made  on adaptation and its relation with
 sensitivity. Especially it is important to 
explain  why the adaptation is always
perfect, i.e. the activity  always returns to the  pre-stimulus level 
 precisely \cite{perfect}; recent experiments  
showed that this perfectness  of  adaptation is robust though  other
properties,  such as the time needed to complete the
adaptation,   vary  with
 conditions \cite{leibler}.

In this letter,  to improve 
  the  previous   approach,  we 
 present  a so-called adaptive Ising model (AIM), in which
 there is a negative feedback from the magnetization on the field.
{\em With large separation of  time
scales},   there exists quasi-equilibrium, which is temporally local, i.e.
on   a short  time scale. On a long time scale, however,
the system evolves, with the existence of 
a dynamical  attractor corresponding   to 
a fixed  pre-stimulus activity and is sensitive  to the  subsequent stimulus.
Thus  we   obtain    a natural  explanation for  why the  
adaptation is always perfect,
 and  show  that not only the sensitive
signal, but also the  effective 
adaptation,  is very likely  a manifestation  of 
 the coupling among receptor dimers. 
Therefore   sensitivity and  adaptation are two sides of the same coin.
Our model is above the level of molecular details, hence it is generic for
signalling networks with similar conditions. Furthermore, the
feedback from the information 
output on the input is a  general way to preserve
sensitivity. 

Under the assumption of high gain limit,  the state of
the receptor dimer, chracterized  as
 $V_{i}$, can be either of two values, $V^0$ and  $V^1$, 
corresponding to the higher and
lower rates of CheA autophosphorylation, respectively.
As in the  neural network \cite{hopfield},
we  assume  McCulloch-Pitts behavior for 
 $V_{i}$, i.e.,  in the absence of noise, 
$V_{i}$$=$$\psi(\sum_{j}T_{ij}V_{j}+H_{i}-U_{i})$,
where 
$\psi(x)=V^1$ if $x>0$  while
$\psi(x)=V^0$ if $x\leq 0$, 
  $U_{i}$ is a threshold value, $T_{ij}$ describes  coupling between 
neighbouring receptor dimers, $H_{i}$ is the effect of ligand binding and 
 methylation level change.
With $T_{ij}=T_{ji}$ and $T_{ii}=0$,  there exists 
a  Hamiltonian which determines the equilibrium state.
One may use  the spin representation $S_{i}=2(V_{i}-V^0)/\Delta V-1$,
where $\Delta V=V^1-V^0$.
 Defining  $J=J_{ij}=T_{ij}\Delta V^2/4$ 
and $B_{i}=H_{i}\Delta V/2$,  
assuming that the ``magnetization'' is  zero 
at the  paramagnetic phase for  $c=0$, and taking into account 
the time-dependence of  $B_{i}$,
we obtain the    Hamiltonian 
\begin{equation}
{\cal H}(t)\,=\,-\sum_{\langle ij\rangle} J_{ij} S_{i}S_{j}-\sum_{i}B_{i}(t)S_{i},
\label{ising1}\end{equation}
where  $\langle ij\rangle$ denotes nearest neighbouring pairs. 
 $J_{ij}=J> 0$ is a constant. 
An essential element of  AIM  is a negative feedback on $B_{i}$:
\begin{equation}
\frac{dB_{i}(t)}{dt}\,=\,-\sigma S_{i}(t-t_r), \label{cb1}
\end{equation}
where  $\sigma>0$, $t_r$ is the retard time of feedback.
This is not an arbitrary  
assumption put in by hand, but is a close 
representation of the experimental finding  that,
in the chemotactic signalling pathway,  
 the state  of CheA, through CheB and CheR, causes an opposite
 effect on its  state later on  \cite{stock}.  
 The initial 
condition is  that 
$B_{i}(t_0)$ is bimodally distributed between $B$ and $0$,
\begin{equation}
p[B_{i}(t_0)]\,=\, c\delta[B_{i}(t=t_0)-B]+(1-c)\delta[B_{i}(t_0)].\label{dis0}
\end{equation}   
Here  $B_i=B$ if the receptor dimer
$i$ is bound to ligand, otherwise $B_i=0$. $B>0$ for attractant binding while
$B<0$ for repellent binding.
$c$ is the  the fraction
of the receptor dimers with ligand bound,
determined by the ambient concentration of 
the chemical.
In other words,   $B_i(t)$  is superposed by two
parts.  One part is the  externally applied  field
 $B_{i}(t_0)\theta (t-t_0)$, where 
$\theta(x) =0$ 
if $x<0$ while 
$\theta(x)=1$ if $x\geq 0$.
Another part  is an induced field, denoted as  $M_{i}(t)$,
with 
$dM_i(t)/dt=-\sigma S_{i}(t-t_r)$.
Note that we have assumed that the randomness  is quenched, representing that 
ligand binding is so strong that debinding happens on a very 
long time scale \cite{bind}. 

Generally, AIM defines a non-equilibrium model. However,  
the large  separation of time scales
 holds in the current problem: ligand binding and conformation change occur  within
only   millisecond,  demethylation reactions take about $0.1$
seconds, 
while time needed to complete the adaptation, which is  associated with the 
slow modulation of methylation level,
is on  the scale of   
 minutes \cite{stock}. 
This situation validates an adiabatic approximation.
Roughly speaking, the  demethylation reaction time plus the time
for the transfer of phosphorylation group from CheA to CheB 
corresponds to the retard time of feedback $t_r$.
Therefore, compared with  
the time needed to achieve temporally local equilibrium,
the evolution  of $B_i(t)$ is very slow, i.e.
the time scale of the  adaptation  process, characterized by
$t_r$,  is very long,
Hence  we may solve our problem by using coarse-graining,
replacing  the above Hamiltonian with a temporally 
coarse grained one 
\begin{equation}
{\cal H}(\tau)\,=\,-\sum_{\langle ij\rangle } J_{ij} S_{i}S_{j}-\sum_{i}B_{i}(\tau)S_{i},
\label{ising}\end{equation}
where $\tau$  is
the  coarse grained and discretized time defined as $\tau= int(t/t_r)$.
Here the function $int(x)$ is the greatest 
integer less than or equal to  $x$.
${\cal H}(\tau)$ determines, through  equilibrium statistical mechanics, 
the coarse grained instantaneous 
state  characterized by the magnetization per 
spin,
$m(\tau)$, which is the value of  each $S_{i}(\tau)$.
Note that $S_{i}(\tau)=m(\tau)$  is the average  of  $S_i(t)$ over the time 
period from $(\tau-1)T$
to  $\tau T$,
equal to the ensemble  average $\langle S_{i}\rangle(\tau)$.
On the long time scale, $m(\tau)$ depends on $\tau$  because of the
feedback on $B_{i}$.
$m(\tau)$ is a measure of the signal. Usually the signal is characterized
by the change of the number of the receptor dimers with state
$V^0$, the average of which is $m/2$.

On the  coarse grained time scale,
the initial condition becomes that
$B_{i}(\tau_0)$ bimodally distributes 
between $0$ and
$B$, i.e.
$p[B_{i}(\tau_0)]$$=$$c\delta[B_{i}(\tau=\tau_0)-B]
+(1-c)\delta[B_{i}(\tau_0)]$,
where $\tau_0 = int(t_0/t_r)$.
The feedback equation becomes
$B_{i}(\tau)=B_{i}(\tau-1)-\sigma m(\tau-1)$,
or equivalently,
$M(\tau)=M(\tau-1)-\sigma m(\tau-1)$,
which implies
\begin{equation}
M(\tau)\,=\,M(\tau_0) -\sigma \sum_{\tau_0}^{\tau -1} m(k) \label{mtau}.
\end{equation}
On the coarse grained time scale, the induced field is the same 
for different spins, therefore the subscript $i$ has been omitted.

Under the adiabatic approximation, we apply mean field theory for 
each instant $\tau$ to obtain:
\begin{equation}
m(\tau)=
\frac{2c}{1+\exp[-2\beta(\nu Jm(\tau)+M(\tau)+ B)]}
+\frac{2(1-c)}{1+\exp [-2\beta(\nu Jm(\tau)+ M(\tau))]}-1,
\label{mag0}
\end{equation}
where $M(\tau)$ is given by (\ref{mtau}),
  $\beta=1/k_{B}T$, $\nu$ is the number of nearest neighbors.  

One may observe  
that $m=0$ is a fixed point of 
Eq. (\ref{mag0}):
if $m(\tau-1)=0$, then $m(\tau)=m(\tau-1)=0$. Moreover,
if 
 $\sigma/\nu J$ is small enough,
 $m(\tau)$ does not change the sign while  its
magnitude decreases towards $0$ \cite{exp0}.
Therefore $m=0$ is an attractor of the evolution of the magnetization.

In the original Ising model with $c=0$,  there are two phases, ferromagnetic
 and paramagnetic,  depending on $\beta\nu J$. For AIM, however,
as an interesting consequence of the feedback, $m(c=0)$ is always zero:
 suppose  $m(c=0)$ is nonzero initially, the feedback  
automatically causes it to attenuate to zero. 
Therefore, we always have $m(\tau < \tau_{0})=0$, and thus
$M(\tau_0)=0$.  Consequently
\begin{eqnarray}
m(\tau \geq \tau_0 )=&
\frac{2c}{1+\exp[-2\beta (\nu Jm(\tau)-\theta(\tau-\tau_0-1)
\sigma\sum_{k=\tau_0}^{\tau-1}
 m(k)+ B)]} \nonumber \\
&+\frac{2(1-c)}{1+\exp [-2\beta(\nu Jm(\tau)-\theta(\tau-\tau_0-1)
\sigma\sum_{k=\tau_0}^{\tau-1}
 m(k))]}-1.
\label{mag}
\end{eqnarray}

Thus  when a ``field'' is  applied, i.e. ligands are bound 
to the receptor dimers, randomly but
 with a certain  occupancy $c$, 
 there is an initial change of magnetization
from $0$ to $m(\tau_0)$,  
depending   on  $c$. This 
initial response   can be arbitrarily sensitive, as seen from
$\partial m(\tau_0)/\partial c$ with $c\rightarrow 0$,
 given by Eq. (10) of Ref. \cite{shi}. 
However, due to the negative feedback of the output (magnetization)
on  the input (field) at  each spin, the magnetization
always attenuates towards zero. Practically, 
 the adaptation is   completed 
 when the difference between  $m(\tau)$ and 
zero is below the detectable threshold of the motors.

Note that in Ref. \cite{shi}, it  had to set that  $\beta\nu J\leq 1$, i.e.
  the system should be  in the paramagnetic 
phase when there is no ligand binding.
 With the feedback naturally integrated to the model in an {\it ab initial}
way, this constraint becomes unnecessary, and thus the model becomes more
robust.

To obtain  some analytical  sense, consider  high temperature limit
$\beta\rightarrow 0$. In this case,
 $m(\tau_0)=c\beta B/(1-\beta\nu J)$ \cite{shi}.
 A simple calculation  based on
Eq. (\ref{mag}) reveals that $m(\tau_0+\Delta \tau)
=[1-\beta\sigma/(1-\beta\nu J)]^{\Delta \tau}m(\tau_0)$.
When $\beta\sigma< 1-\beta\nu J$, $m(\tau)$ attenuates towards zero
exponentially.
For generic   values of the parameters, the solution  can only be obtained
numerically, as shown in Fig. 2. Note that the effective parameters
are $\beta\nu J$, $\beta B$, $\beta\sigma$, and $c$.
  Comparing plots for different values of
parameters, one  can observe  that the speed of attenuation of $m(\tau)$ 
increases with  $\beta\sigma$ and with $\beta\nu J$, while 
 decreases with $c$.
It increases with $\beta B$, but  when $\beta B$ is large enough,
$m(\tau)$ becomes  independent of the exact value of $\beta B$, as indicated 
by the results for $\beta B=1,10$ with $\beta\nu J=0.5$. On a 
log-log scale  (not shown), the plots are generally convex,
indicating that the attenuation is in general  
more rapid than exponential decay, due to the larger $\beta\nu J$.

After the adaptation is completed, if there is
a further change in the chemoeffector 
concentration, thus the occupancy changes from $c$ to $c+c'$ at $\tau_0'$,
 then $m(t\geq \tau_0')$ is  given by Eq. (\ref{mag0})
with $c$ updated with $c+c'$. Because $m(\tau_0'-1)=0$,
 $M(\tau_0'-1)$ is given by 
\begin{equation}
0=\frac{2c}{1+\exp[-2\beta (M(\tau_0'-1)+ B)]} 
+\frac{2(1-c)}{1+\exp [-2\beta M(\tau_0'-1)]}-1.
\label{mag00}
\end{equation}
Hence,
\begin{eqnarray}
m(\tau \geq \tau_0')=&
\frac{2(c+c')}{1+\exp[-2\beta(\nu Jm(\tau)-
\sigma\sum_{k=\tau_0'}^{\tau-1} m(k)+M(\tau_0'-1)+B)]} \nonumber \\
&+\frac{2(1-c-c')}{1+\exp [-2\beta(\nu Jm(\tau)-
\sigma\sum_{k=\tau_0'}^{\tau-1}m(k)+M(\tau_0'-1)]}-1
\label{magn}
\end{eqnarray}
which is largely determined by $c'$ since the effect of $c$ is counteracted
by $M(\tau_0')$. $m(\tau > \tau_0')$
attenuates towards zero, repeating  the  dynamics of 
 Eq. (\ref{mag}).
 $\partial m(\tau_0')/\partial c'$ with $c'\rightarrow 0$, approximately 
equal to $\partial m(\tau_0)/\partial c$ with $c\rightarrow 0$,  can be
 arbitrarily large if the latter can.
Therefore our adaptation mechanism not only
brings the signal to the pre-stimulus level, but also 
preserves the sensitivity, as required by chemotaxis.

Therefore  we have explained  
 why perfect adaptation can always be achieved in chemotaxis:
a fixed  pre-stimulus activity
 is a dynamical attractor.
 The  variation of the values of the parameters, under a  basic requirement
that  $\sigma$ is  sufficiently small,  only affect
the time needed to achieve perfect  adaptation. Thus our result is fully
consistent with the experiments.

Recent experimental  analyses of the aspartate receptor revealed
that attractant binding induces a displacement of one of four helices,
each two of which constitute a subunit
 of a receptor dimer \cite{chevitz}.   
Therefore $V_i$ may be identified as the position of the mobile helix 
\cite{exp}.
$V^0$ is the original position of the helix, corresponding to the
higher rate of CheA autophosphorylation.
 $V^1$ is down towards  the cytoplasm, corresponding to the
lower rate of CheA autophosphorylation.
Thus $H$  is the force generated by  ligand binding.
 $2B = H \Delta V$ is the shift of energy difference between the two 
conformations due to free energy exchange with the bound ligand,
or the work done by the generated force. One may find  that
$4J/\Delta V$ is the force due to the activity change of one
nearest neighbour. $2M_i(t)/\Delta V$ is 
the force due to feedback, and thus should be
opposite to the force 
generated by ligand binding.

In the high temperature limit,  when 
$\Delta \tau=-ln2/ln[1-\beta\sigma/(1-\beta\nu J)]$,
 $m(\tau_0+\Delta \tau)=m(\tau_0)/2$.
Assuming  $1/\beta\approx 4pN\cdot nm$,
$\beta\nu J\approx0.5$ \cite{shi}, and that
 the time needed to complete adaptation
be  $1$ minute, i.e.  $\Delta \tau \approx 600$,
 we may estimate that
$\sigma\approx 0.002pN\cdot nm$. 
Because   the formula  is for high temperature limit,
the real value of $\sigma$ is  smaller for the
  assumed values of the  parameter values.
  Experimentally, by measuring $\beta$, 
$\nu J$, $B$ and the  adaptation  time, $\sigma$ can be determined. 
On the other hand,
$\sigma$ can also be determined through 
$\sigma=-[M_i(t)-M_i(t_0)]/\int_{t_0}^{t} S_i(t-t_r) dt$ 
$=[M(\tau_0)-M(\tau)]/\sum_{\tau'=\tau_0}^{\tau-1} m(\tau')$.  
By comparing the results obtained in  different ways, the model
may be tested or refined.

Eq. (\ref{cb1}) implies  that the feedback is assumed to be local.
This is because we preserve the assumption that there exists a 
 feedback loop for each receptor  dimer 
 although we consider  coupling  between the states of neighbouring 
 dimers.
However, one may  make a straightforward 
 extension to  include  the
neighbouring states in the feedback equation, without  
changing 
 the qualitative physics. Furthermore, this makes no change in
the temporally 
coarse grained   feedback equation, or in the sense of (spatial) 
renormalization group. 
Therefore the large separation of time scales, which validates coarse graining,
makes the essential mechanism not so much dependent on the microscopic 
 details. This is also an aspect of robustness.

Finally, it is interesting to note that in the case of bacterial chemotaxis,
the clustering of receptors exists  prior to  a stimulus, 
while in many other cell signalling  processes,
the receptors diffuse on the membrane and the  clustering appears as a 
response to the stimulus. 
One may make a generalization of our 
model to a sort of combination of Ising and lattice gas models to address
such a case, as  will be described in a forthcoming  paper. 
  
To summarize,  an adaptive Ising model is proposed to combine 
cooperativity and negative feedback. It is applied  to
the receptor network  of  
bacterial chemotactic signalling and  explains
 the perfect adaptation as a dynamical 
attractor.
{\em Both the signal magnitude  and the sensitivity of response
are adapted}.
 The large separation of time scales   leads up  to the solution by 
using  adiabatic approximation.   The change of  parameter values,
under a  basic requirement that the feedback effect 
is sufficiently  weak, only changes  the time needed to 
complete  adaptation,
 without affecting   its  perfectness. Hence  the
robustness of perfect adaptation is explained. 
This  work shows that coupling among receptor dimers gives rise to 
a unifying description of  both sensitivity and  effective adaptation. 
We anticipate further experimental and theoretical investigations.
Cooperativity in cell  signalling is likely a new playground of
statistical mechanics.
Combining cooperativity and feedback,  and preserving sensitivity,
 the idea of  AIM may 
be useful for a variety of  problems.

I thank Tom Duke and  Dennis Bray  for discussions.

\begin{figure}
\begin{figure}
\caption{A schematic illustration of the  chemotactic signalling pathway of
one receptor dimer. Around 2000 receptor dimers constitute a network.}
\end{figure}

\caption{Attenuation of $m(\tau)$, the solution of 
Eq. (\ref{mag}), for different values of parameters.
$\tau$ is the coarse grained time, $\tau_0$ is set  to $1$.
To compare the attenuation   speed for different values of 
parameters, we plot
$m(\tau)/m(\tau_0)$. The parameters $(\beta\nu J,\beta B,\beta \sigma,c)$
for each plot are given on the right upside.}
\end{figure}


\begin{references}
\bibitem{perfect} We believe that  
  the perfect adaptation  is necessary for 
chemotactic machinery  to work, as 
an example of perfectness  required 
in many  biological processes.
Suppose that the adaptation is not perfect, and as a generic case, 
 there is  a  probability  distribution  for the  difference 
between  the  activities  after two consecutive  adaptations,
with  mean $\delta$ and  variance $\epsilon^2$.
Then according to the central limit theorem,
after $n$ times of stimulus  and adaptation, 
the activity drifts from the first  pre-stimulus one
with mean $n\delta$ and variance
$n\epsilon^2$.
In the life of the bacterium, $n\rightarrow \infty$, therefore 
the activity range of the  chemotactic machinery should be infinity, 
unless both $\delta$ and $\epsilon^2$ are exactly zero,  
i.e. the adaptation is perfect.
The viewpoint of evolution may thus  help us to understand why 
the  adaptation has to   be  perfect.
On the other hand,   an underlying   physical
mechanism needs to be found.
\bibitem{bind} The time scale of ligand debinding  
 ranges from tens of seconds to tens of minutes, 
longer than the signalling time scale, i.e. the time scale for achieving 
the  
temporally local equilibrium, as discussed below.  This quenched 
disorder model is thus more favorable than a grand canonical 
ensemble approach, which would
treat binding-debinding processes as a part of the process towards 
an  equilibrium.  Although we are not sure whether 
the debinding time scale is always longer than the time  to complete
the adaptation,  this does not  matter, because on the coarse
grained time scale, Eq. (\ref{mag0}) 
is always valid, independent on which of the
receptors are liganded.
\bibitem{exp0} A sufficient condition is $\sigma/\nu J <|x_0|/|m_0|$,
where $x_0$ is the solution to 
$0=$$\frac{2c}{1+\exp[-2\beta(\nu Jx+ B)]}$$+
\frac{2(1-c)}{1+\exp [-2\beta\nu Jx]}-1$.
To understand this, consider $m(\tau-1)$ as the 
$x$ coordinate of the cross between $y=x$ and $y=f(x)$, 
while $m(\tau)$ as the $x$ coordinate of the  cross between $y=x$  and 
$y=f[x-\frac{\sigma}{\nu J}m(\tau-1)]$, which is obtained by translating
$y=f(x)$ along $x$ direction.
\bibitem{exp} We conjecture that
which subunit provides the mobile helix may be  random, and 
  that  dimerization of receptors might
  provide a redundancy so that if  one
subunit is damaged, the other can work as an alternative. 
The negative cooperativity between the two subunits may be due to an
``antiferromagnetic'' coupling.
\end{references}
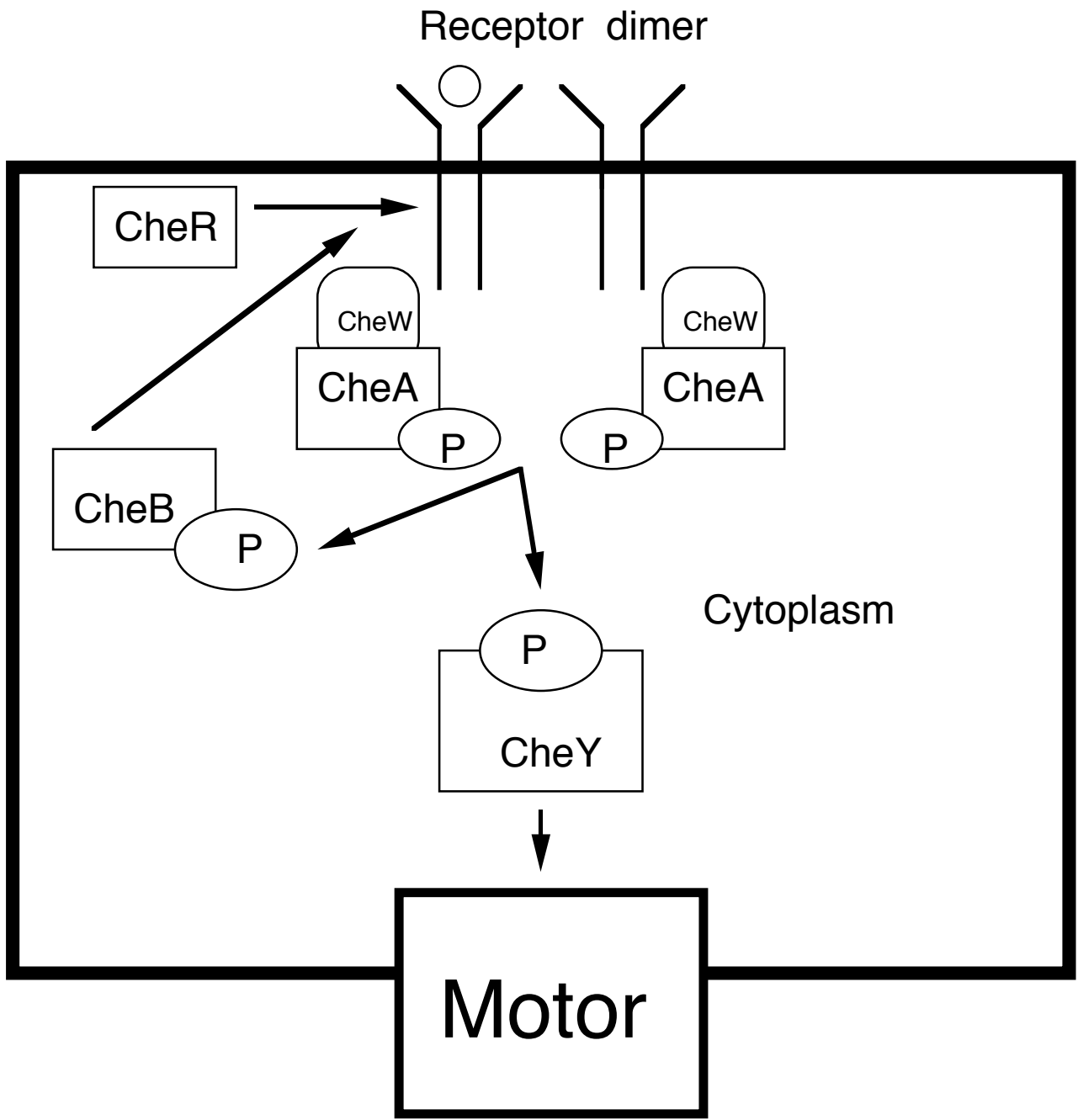
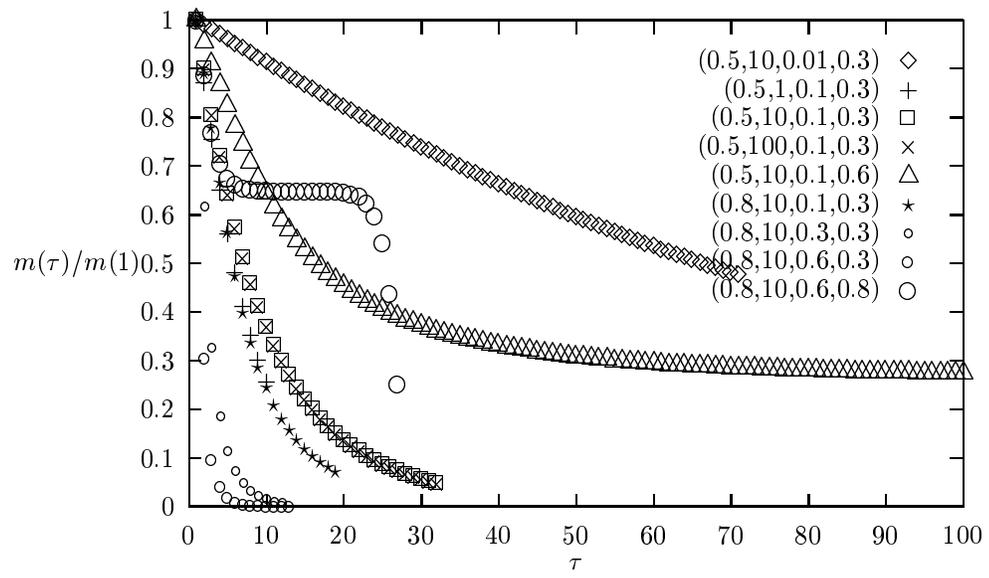
\end{document}